\documentclass{aa}
\usepackage{graphicx}
\begin{document}
\title{VLA detection of OH absorption from the elliptical galaxy \\
       NGC 1052}

\author{Amitesh Omar\inst{1}
\and
K.R. Anantharamaiah\inst{1}\thanks{Deceased since Oct. 29, 2001}
\and
Michael Rupen\inst{2}\
\and
Jane Rigby\inst{3}
 }

\offprints{Amitesh Omar}

\institute{Raman Research Institute, C.V. Raman Avenue, Bangalore, 560 080, India. \\
\email{aomar@rri.res.in}
\and
National Radio Astronomy Observatory, Socorro, NM, USA. \\ 
\email{mrupen@aoc.nrao.edu}
\and
Steward Observatory, University of Arizona, 933 N. Cherry Ave, Tucson, AZ 85721. \\ 
\email{jrigby@as.arizona.edu}
		}
\authorrunning{Omar et al.}
\titlerunning{OH absorption from NGC 1052}
   \date{Received .. 2001 / Accepted .. 2001}

\abstract{
   VLA observations of OH absorption towards the elliptical galaxy NGC 1052
are presented. Both OH lines, at 1665 and 1667 MHz, were detected in
absorption towards the center of NGC 1052. The hyperfine ratio of the two OH
lines ($\tau$$_{1667}$/$\tau$$_{1665}$) is 2.6$\pm$0.8 as compared to 1.8
expected for the excitation under LTE conditions for an optically thin
cloud.  The column density of OH is estimated to be
$2.73(\pm0.26)\times10^{14}~cm^{-2}$ assuming $T_{ex} \sim$10 K.  The
centers of both the OH lines are redshifted from the systemic velocity of
the galaxy by $\sim$173 km s$^{-1}$. The velocity of OH line coincides with
the velocity corresponding to the strongest HI absorption. We suggest that
OH absorption is arising from a molecular cloud falling towards the nucleus.
The OH line, though narrower, is found to be within the much broader and
smoother H$_2$O megamaser emission. The possible link between OH/HI and
$H_2O$ emission is discussed. 
  } 
\maketitle
%________________________________________________________________

\section{INTRODUCTION} 
The most extensive and conclusive confirmation for the presence of cold
interstellar material in early-type galaxies came from observations of dust
with the Infrared Astronomical Sattelite (IRAS) (\cite{neu84, kna85,
kna89}).  Sensitive observations of HI (\cite{van89,huc95}) have also shown
that elliptical galaxies contain a significant amount of cold interstellar
matter. The molecular contents of elliptical galaxies has been studied
mainly through CO observations of infrared bright elliptical galaxies
(\cite{wan92,wik95,kna96}). These observations resulted in the detection of
molecular gas in several galaxies in emission and four galaxies in
absorption, indicating that the overall detection rate of CO in elliptical
galaxies is about 10--15\%. The OH radical in absorption is also a good
tracer of molecular gas in interstellar clouds (\cite{lis96}). Single dish
OH surveys (\cite{sch86, baa92, sta92, dar00}) of several hundred galaxies
of various types resulted in the detection of about 3 dozen galaxies, of
which none was an elliptical.

NGC 1052, a moderately luminous
(L$_{b}=1.6\times10^{10}$L$_{\sun}$) elliptical
galaxy of type E4, is a member of a small group in the Cetus--I cloud. There
are several estimates of the velocity for this system in the literature,
which differ from each other by a few tens of km s$^{-1}$. We adopt
V$_{hel}=1474\pm10$ km s$^{-1}$, estimated from the optical emission lines
(\cite{dev91}), which implies that NGC 1052 is at a distance of 21 Mpc
(assuming H$_0=70$ km s$^{-1}$ Mpc$^{-1}$ and q$_0=0$).  It is classified as
a LINER (\cite{fos78, ho97}) and is known for its several water megamasers
(\cite{bra96},~\cite{cla98}).  HI absorptions, redshifted from the systemic
velocity, were detected at 1486, 1523 and 1646 km s$^{-1}$ against the
nuclear continuum source (\cite{van86}). NGC 1052 was reported to have CO
emission as well as absorption by \cite{wan92}, but later observations by
\cite{wik95} failed to confirm those detections. More recently, \cite{kna96}
have reported a possible CO absorption from NGC 1052 near 1622 km s$^{-1}$.  
Since the reported CO detections are quite noisy, it remains uncertain
whether NGC 1052 has a molecular component associated with the HI (21cm)
absorption.

Here we report the first detection of 1665 and 1667 MHz OH absorption in NGC
1052.  The next section describes the observational details and results.
Subsequent sections compare these results with observations at optical,
X-ray, and other wave bands, and discuss some of the implications.

%__________________________________________________________________

\section{OBSERVATIONS \& RESULTS} 
NGC 1052 was observed in the B configuration of the VLA, which has
interferometric baselines ranging from 100 m to 11 km. Data were recorded in
the 4IF correlator mode, recording 1.5625 MHz in each of the two circular
polarizations for two frequency bands, one centered at 1656.5 and other at
1658.3 MHz. The details of the observations are listed in
Table~\ref{tab:obspar}. The data were reduced in AIPS using standard
calibration and imaging methods. The amplitude, phase and frequency response
of the antennas were calibrated separately for each IF.  The phase and
amplitude gains of the antennas were derived from observations of the
standard VLA calibrator 0240--231 at intervals of 30 minutes.  The flux
scale was set using \cite{baa77} flux density of the standard VLA calibrator
3C48.  A combined bandpass spectrum was generated using all the data taken
on the amplitude and phase calibrators as well as on the strong radio source
0319+415 (3C84). A continuum data set was formed by averaging the calibrated
visibility data of 50 line-free channels. The continuum data set was
self-calibrated and the resulting antenna gain corrections were applied to
every spectral channel separately.  The continuum emission common to all
channels was removed using the task `UVLIN' inside AIPS. Continuum--free
images for all channels were made and the source region was searched for
absorption.  Both 1665 and 1667 MHz lines were detected, in each of the two
circular polarizations. Although, a part of the band centered at 1656.5 MHz
was affected by interference, the detected 1665 MHz line was outside the
affected region.

\begin{table}
\caption{\bf{Observation Parameters}}
\label{tab:obspar}
\begin{tabular}{ll}
\hline
\hline
\bf{Parameter} & \bf{Value} \\
\hline
Date of Observation &1998 Sep03 \\
RA, Dec (J2000.0) &02 41 04.79,--08 15 20.75 \\
%Declination (J2000.0) &--08 15 20.75 \\
%VLA configuration &B \\
Observing duration (hrs) &5 \\
Range of baselines (km) &0.1--11 (B config)\\
Observing frequencies (MHz)(IF1,IF2) &1656.50, 1658.30 \\
%Observing frequency (MHz)(IF2) &1658.30 \\
Bandwidth per IF (MHz) &1.562 \\
Number of spectral channels &64 \\
Polarizations &RCP \& LCP\\
Synthesised beam (Natural Weight) &$6.4^{''}\times4.3^{''}$, PA$=9.7\degr$ \\
Velocity resolution &4.4 km s$^{-1}$ \\
Frequency resolution (kHz/channel) &24.4 \\
%Hanning smoothing &offline \\
Amplitude calibrator &0137+331 (3C48) \\
Phase calibrator &0240--231 \\
Bandpass calibrator &0319+415 (3C84) \\
rms noise per channel (mJy beam$^{-1}$) &0.7 \\
\hline
\end{tabular}
\end{table}

\begin{figure}
\centering
\includegraphics[height=7.5cm,width=7.5cm,angle=0]{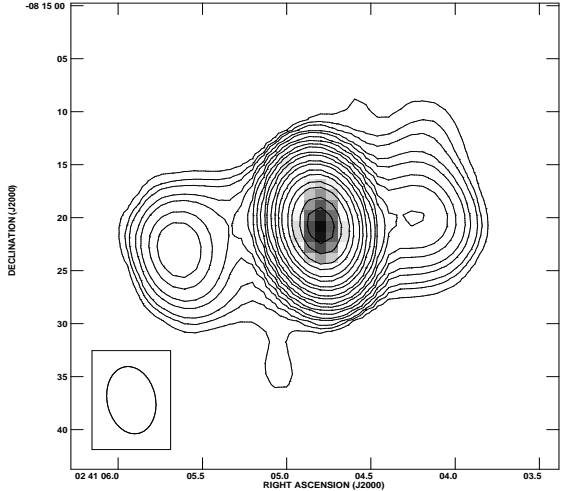}
\caption{The radio continuum image of NGC 1052 drawn as contours with levels
of 1.8 mJy beam$^{-1}\times$
(1,1.5,2,3,4,6,8,12,16,24,32,48,64,96,128,192,256,384,512). The peak flux
density in the contour image is 1.14 Jy beam$^{-1}$. The peak flux densities
of the E and W lobes are 22.3 and 19.4 mJy beam$^{-1}$ respectively.  The
grey scale represents the velocity--integrated optical depth of the 1667 MHz
OH absorption. The synthesised beam depicted in the bottem left corner is
$6.4^{''}\times4.3^{''}$, PA$=9.7\degr$. } 
\label{fig:cont} 
\end{figure}

The core/jet morphology in the continuum image of NGC 1052 is in accordance
with the previous observations by Jones et al. (1984). The peak continuum
flux density of the core is $\sim$ 1.14 Jy. The total flux density including
contributions from the two radio lobes is $\sim$ 1.23 Jy. The continuum
image (Figure~\ref{fig:cont}) shows that the radio axis is at a position
angle (E to N) of 103\degr. The two radio lobes are asymmetrically located
about the radio nucleus, being 14$^{''}$ to the east and 8$^{''}$ to the
west. The continuum nucleus and the line absorption are unresolved with the
synthesised beam ($6.4^{''}\times4.3^{''}$, PA$=9.7\degr$). Both 1665 and
1667 MHz lines are detected at a redshifted velocity of $\sim$173 km
s$^{-1}$ with respect to the systemic velocity of the galaxy.  The column
density of OH can be estimated from

\begin{equation}
\label{eqn:colden}
N_{OH} = 2.35 \times 10^{14} T_{ex}~\int~\tau_{1667}~dV ~cm^{-2}
\end{equation}

\noindent (\cite{dic81,lis96}) where T$_{ex}$ is the excitation temperature
in Kelvins, $\tau_{1667}$ is the optical depth of the 1667 MHz line and V is
the velocity in km s$^{-1}$; for NGC 1052, above equation gives an OH column
density of $2.73(\pm0.26)\times10^{14}$~(T$_{ex}/10$) ~cm$^{-2}$ towards the
center. For the two lobes, we estimate an average 3$\sigma$ upper limit of
OH absorption as $\sim$ 0.10. This upper limit implies that 0.6\% absorption
seen towards the nucleus is undetectable from either of the lobes even if
absorbing gas covers the entire continuum source.

The AIPS gaussian fitting routine `SLFIT' was used to derive the line
parameters. The peak optical depth of the 1667 MHz line is
$5.8(\pm0.2)\times10^{-3}$ and that of the 1665 MHz line is
$2.9(\pm0.1)\times10^{-3}$. The FWHM of 1667 and 1665 MHz lines are
18.8$\pm$1.3 and 14.5$\pm$2.6 km s$^{-1}$ respectively. Given the
uncertainity in the overall shape of the 1665 MHz line due to low optical
depth, profiles of the 1665 and 1667 MHz lines can be considered similar.  
The ratio of the integrated optical depth is 2.6$\pm$0.8 which is marginally
higher than that expected (viz.  1--1.8) for excitation in thermal
equillibrium. The mean value of 1667 to 1665 MHz line ratio is about 1.6 for
galactic diffuse clouds (\cite{dic81}).

\section{DISCUSSION}
\subsection{Link with HI and X-ray absorbing column}

\begin{figure}
\centering
\includegraphics[height=8cm,width=7.0cm,angle=0]{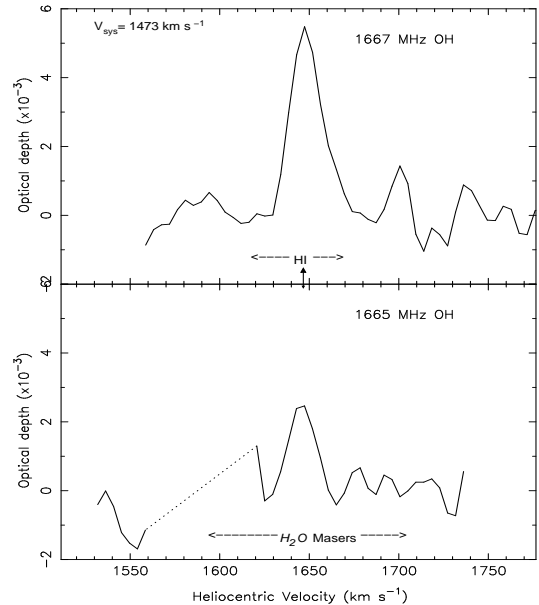}
\caption{A plot of the optical depth of 1667 and 1665 MHz absorption lines
towards the nucleus of NGC 1052. The spectrum has been Hanning smoothed
offline using a window of 3 adjacent channels. The figure displays the
entire velocity coverage by VLA observations. The region marked by dashed
lines in 1665 MHz spectrum was affected by interference. The velocity range
over which HI absorption and H$_2$O masers are observed are indicated in the
top and bottom frames respectively. The systemic velocity is indicated on
top left corner of the upper frame.}
\label{fig2}
\end{figure}
 
HI components in NGC 1052 have been seen in absorption at 1486, 1523 and
1646 km s$^{-1}$, which are redshifted from the systemic velocity
(\cite{van86}). The N(HI)/T$_{ex}$ values of three components are
$0.6\times10^{18}$, $1.0\times10^{18}$ and $1.4\times10^{18}$ cm$^{-2}$
respectively. The strongest absorption ($\tau \sim0.02$) is at 1646 km
s$^{-1}$ with a FWHM of about 35 km s$^{-1}$. Due to the similarity in the
velocity of OH absorption with the highest redshifted component of HI
absorption, it is reasonable to associate this HI component with the OH
detected in these observations. It is interesting that the velocity of OH
absorption matches very well with the strongest HI absorption component at
1646 km s$^{-1}$ even after a difference of about 16 years in the
observations. The stability of OH/HI line suggests that the absorbing cloud
covers a substantial fraction of the milliarcsec VLBI core in which most of
the radio emission lies (\cite{jon84, kam01}). The integrated optical depth
ratio of HI to OH is $\sim6$, which is in accordance with the values
obtained for the galactic diffuse clouds (\cite{dic81}). The linewidth ratio
of HI to OH is $\sim2$, which suggests that the excitation of OH is
restricted to some preferred regions inside the cloud. If redshifted
absorption is considered as an evidence of infall of gas to the nucleus,
where a small fraction of the gaseous mass is converted to luminosity, then,
the association of a large amount of molecular gas with the neutral gas will
imply a lower efficiency of the central engine in converting mass to
luminosity. The observed line widths (FWHM) viz. $\sim18$ km s$^{-1}$ of the
two OH absorption is considerably higher than would be expected ($\sim1$ km
s$^{-1}$) from purely thermal motions, assuming the gas temperature is at
most a few tens of K. However, if the gas is very close (within few pc) to
the nucleus, some kinematical effects will tend to broaden the observed
absorption line e.g., turbulence may set up to overcome the gravitational
collapse against the nucleus. If the gas is in a disk, then, a velocity
gradient along the disk, as seen in some megamaser galaxies (e.g.
\cite{hag00}), can explain the observed line width of the OH absorption. On
the other hand, if the observed dispersion is considered due to conglomerate
of individual clouds in virial equilibrium, a binding mass will be about
10$^{6}$ M$_{\sun}$, a value close to that seen in some giant molecular
clouds
(GMCs) of our galaxy. The typical velocity width of such GMCs has been
estimated close to 10~km s$^{-1}$ (\cite{sta78}).

	The gas is expected to be much hotter in the vicinity of an AGN due
to enhanced Ly$\alpha$ pumping which in turn will increase the T$_{ex}$ to a
few thousand kelvin.  Assuming, T$_{ex}\sim1000$~K, the predicted total
N(HI) will be $2.0\times10^{21}$ cm$^{-2}$ including all three HI
components.  For the detected OH component, taking the relative abundance
ratio of OH/H$_2$ $=1\times10^{-7}$ (\cite{gue85,lis99}), the implied column
density of H$_2$ is $2.73\times 10^{21}$~(T$_{ex}/10$) ~cm$^{-2}$. The
implied CO column density is about $5.5\times 10^{14}$~cm$^{-2}$, which is
about 10 times higher than predicted from CO observations. In comparison,
X-ray observations indicate a hydrogen column density greater than
$1\times10^{23}$ cm$^{-2}$ (\cite{wea99}), which is significantly higher
than the total hydrogen column estimated via radio observations (HI \& OH).  
This excess column density inferred from X-ray data has been seen in many
active galaxies, and, was explained due to excess absorption by a
combination of dust and partially ionized gas (\cite{gal99}). It should be
noted here that since HI and OH absorptions are spatially unresolved, the
estimated values of OH and HI column densities are only a lower limit.  
Also, X-ray absorption is arising towards the nucleus which is free-free
absorbed at wavelengths corresponding to the HI and OH absorptions
(\cite{kam01}), therefore, radio observations are sampling off nuclear gas
which may be of different composition than the gas probed via X-ray
observations.

\subsection{Link with H$_2$O Megamasers ?}
It is very surprising that the OH absorption, though narrower than the water
maser emission, is coincident with the velocity centroid of the 22 GHz
H$_2$O masers. NGC 1052 is the only known elliptical galaxy having H$_2$O
megamaser emission. The megamasers and their link with AGNs are generally
understood in terms of obscuring torus models. The link is thought to be a
consequence of irradiation of the inner face of the torus by hard X-rays
from the nuclear continuum source, which enhances the water abundance within
a molecular layer at a temperature of 400--1000~K (\cite{neu94}). H$_2$O
megamasers of NGC 1052 are unusual in showing a relatively smooth profile
which moves in velocity over time by about 70 km s$^{-1}$ on a time scale of
a year (\cite{bra96}). Water masers in NGC 1052 are distributed along the
jet rather than perpendicular to it (\cite{cla98}) unlike in NGC 4258 in
which water masers are originating in a torus (see \cite{miy95}). Claussen
et al. (1998) suggested that these masers are excited by shocks in to
circumnuclear molecular cloud, or alternatively, amplifying radio continuum
emission of the jet by foreground molecular clouds. It should be noted that
the shocks can also enhance the abundance of OH by dissociation of H$_2$O
before the gas is cooled down below 50K (\cite{war99}), however, the
observed column density of OH is one order of magnitude less than that
predicted. A drift in the velocity of maser feature was considered as a
consequence of the moving jet which will illuminate different parts of the
foreground H$_2$O masing cloud. Efficient maser emission will take place at
total column density (N$_H$) below the quenching density which is estimated
as 10$^{25}$--10$^{27}$ cm$^{-2}$ for NGC 1052 (see \cite{wea99}). This
upper limit on column density is well above than that predicted from our
observations.  However, it is not clear how HI/OH are quite stable over a
long period of time while H$_2$O emission changes substantially over a short
time scale. Further simultaneous observations of HI, OH and H$_2$O masers
are required to make a connection between molecular gas traced by OH
absorption and H$_2$O masing gas.

\section{SUMMARY}

These VLA observations have resulted in the first detection of OH absorption
in an elliptical galaxy. Both, 1665 and 1667 MHz OH absorption, were
detected from the elliptical galaxy NGC 1052. The linewidths of both the OH
lines are significantly large as compared to that expected for a cloud in
thermal conditions at few tens of K. The gas is predicted to be close to the
nucleus. A remarkable coincidence of velocity is found with the strongest
and redshifted HI absorption and H$_2$O emission, however link to the
megamaser emission is still not understood.  Based on the abundance ratio of
OH/H$_2$ as 1$\times$ 10$^{-7}$, it is predicted that the column density of
molecular gas in NGC 1052 is comparable to HI. Higher angular and spectral
resolution observations would be usefull for detail kinematics of the OH
absorption while simultaneous observations of H$_2$O and HI/OH observations
would be neccessary to understand the link between masing gas and molecular
gas traced by OH absorption.

\begin{acknowledgements}
      The National Radio Astronomy Observatory is a facility of the National Science Foundation
operated under cooperative agreement by Associated Universities, Inc.  
\end{acknowledgements}


\begin{thebibliography}{}
\bibitem[Baan et al. 1992]{baa92} Baan W.A., Haschick, A.D., Henkel, C. 1992, AJ, 103, 728
\bibitem[Baars et al. (1977)]{baa77} Baars, J. W. M., Genzel, R., Pauliny-Toth, I.I.K. \& Witzel, A. 1977, A\&A, 61, 99
\bibitem[Braatz et al. 1996]{bra96} Braatz, J.A., Wilson, A.S. \& Henkel, C. 1996, ApJ, 106, 51
\bibitem[Claussen et al. 1998]{cla98} Claussen, M.J., Diamond, P.J., Braatz, J.A., Wilson, A.S. \& Henkel,
C. 1998, ApJ, 500, L129
\bibitem[Darling \& Giovanelli  2000]{dar00} Darling J. \& Giovanelli, R. 2000, AJ, 119, 3003 
\bibitem[de Vaucouleurs 1991]{dev91} de Vaucouleurs, G.,de Vaucouleurs A., Corwin Jr. et al. In Third Reference Catalogue
of bright galaxies, version 3.9 
\bibitem[Dickey et al. 1981]{dic81} Dickey, J.M., Crovisier, J. \& Kaz\`es, I. 1981, A\&A 98, 271
\bibitem[Fosbury et al. 1978]{fos78} Fosbury, R.A.E., Mebold, U., Goss, W.M. \& Dopita, M.A. 1978, MNRAS, 183,
549
\bibitem[Gallimore et al. 1999]{gal99} Gallimore, J.F., Baum, S.A., O'Dea, C.P., Pedlar, A. \& Brinks, E. 1999, ApJ,
524, 684
\bibitem[Gu\`elin 1985]{gue85} Gu\`elin, M. 1985 In Molecular Astrophysics, ed. by Diercksen, W.F. Huebner, W.F. and
Langhoff, P.W. (Reidel), p.23
\bibitem[Hagiwara et al. 2000]{hag00}Hagiwara, Y., Diamond, P.J., Nakai, N. \& Kawabe 2000, A\&A, 360, 49
\bibitem[Ho et al. 1997]{ho97} Ho, L.C., Filippenko, A.V. \& Sargent, W.L.W. 1997, ApJs, 112, 315
\bibitem[Huchtmeier et al. 1995]{huc95} Huchtmeier, W.K., Sage, L.J. \& Henkel, C. 1995, A\&A, 300, 675 
\bibitem[Jones et al. 1984]{jon84} Jones, D. L., Wrobel, J.M. \& Shaffer, D.B. 1984, ApJ, 276, 480  
\bibitem[Kameno et al. 2001]{kam01} Kameno, S., Sawada-Satoh, S., Inoue, M. , Zhi-Qiang S. \& Kiyoaki, W. 2001, PASJ, 53, 169
\bibitem[Kazes \& Dickey 1985]{kaz85} Kazes, L. \& Dickey, J.M. 1985, A\&A, 152, 9
\bibitem[Knapp et al. 1989]{kna89} Knapp, G.R., Guhathakurta, P. \& Kim, D.W. 1989, ApJS, 70, 329
\bibitem[Knapp \& Rupen (1996)]{kna96} Knapp, G.R. \& Rupen, M.P. 1996, ApJ, 460, 271
\bibitem[Knapp et al. 1985]{kna85} Knapp, G.R., Turner, E.L. \& Cunniffe, P.E. 1985, AJ, 90, 454
\bibitem[Liszt \& Lucas 1996]{lis96} Liszt, H. \& Lucas, R. 1996, A\&A, 314, 917
\bibitem[Liszt \& Lucas 1999]{lis99} Liszt, H. \& Lucas, R. 1999, Highely Redshift Radio Lines, ASP conference series, vol
156, ed. by Carilli et al. p.188
\bibitem[Miyoshi et al. 1995]{miy95} Miyoshi, M., Moran, M., Herrnstein, J. et al. 1995, Nature, 373, 127
\bibitem[Neufeld et al. 1994]{neu94} Neufeld, D.A., Maloney, P.R. \& Conger, S. 1994, ApJ, 436, L127
\bibitem[Neugebauer 1984]{neu84} Neugebauer, G. 1984, ApJ, 278, L1
\bibitem[Schmelz et al. 1986]{sch86} Schmelz, J.T., Baan, W.A., Haschick, A.D. \& Eder, J. 1986, AJ, 92, 1291
\bibitem[Stark \& Blitz 1978]{sta78} Stark A.A. \& Blitz L. 1978, ApJL, 225, L15
\bibitem[Staveley-Smith et al. 1992]{sta92} Staveley-Smith, L., Norris R.P., Chapman, J.M. et al. 1992, MNRAS, 258, 725
\bibitem[van Gorkom et al. 1989]{van89} van Gorkom J.H., Knapp. J.H., Ekers, S.M. et al. 1989, AJ, 97, 708
\bibitem[van Gorkom et al. 1986]{van86} van Gorkom J. H., Knapp, G. R., Raimond E., Faber S. M. \& Gallagher
J. S. 1986, AJ, 91, 791
\bibitem[Wang et al. (1992)]{wan92} Wang Z., Kenney, J. D. P. \& Ishizuki S. 1992, AJ, 104, 2097
\bibitem[Wardle 1999]{war99} Wardle, M. 1999, ApJ, 525, L101.
\bibitem[Weaver et al. 1999]{wea99} Weaver, K.A., Wilson, A.S., Henkel, C. \& Braatz, J.A. 1999, ApJ, 520, 130  
\bibitem[Wiklind et al. (1995)]{wik95} Wiklind, T., Combes, F. \& Henkel, C. 1995, A\&A, 297, 643

\end{thebibliography}
\end{document}